\documentclass[aps,prl,amssymb,amsmath,twocolumn,floatfix]{revtex4}
\usepackage{graphicx}
\usepackage{amsmath}
\usepackage{color}

\usepackage{xr}
\begin{document}
\title{Finite-size analysis in neural network classification of critical phenomena}
\author{Vladislav Chertenkov$^{1,2}$}
\author{Evgeni Burovski$^{2}$}
\author{Lev Shchur$^{1, 2}$}
\affiliation{$^1$ Landau Institute for Theoretical Physics, 142432 Chernogolovka, Russia}
\affiliation{$^2$ HSE University, 101000 Moscow, Russia}

\begin{abstract}
We analyze the problem of supervised learning of ferromagnetic phase
transitions from the statistical physics perspective. We consider two
systems in two universality classes, the two-dimensional Ising model
and two-dimensional Baxter-Wu model, and perform careful finite-size
analysis of the results of the supervised learning of the phases of
each model. We find that the variance of the neural network (NN) output function (VOF)
as a function of temperature has a peak in the critical region.
Qualitatively, the VOF is related to the classification rate of the
NN. We find that the width of the VOF peak displays the finite-size
scaling governed by the correlation length exponent, $\nu$, of the
universality class of the model.
We check this conclusion using several NN architectures---a fully
connected NN, a convolutional NN and several members of the ResNet
family---and discuss the accuracy of the extracted critical exponents
$\nu$.
\end{abstract}

\maketitle

\textit{Introduction.---} Deep learning is since recently emerging as a promising
tool for studying phase transitions and critical phenomena. The pioneering
observation of Ref.~\cite{Carrasquilla2017} is that training a neural network
(NN) to perform a binary classification of microscopic spin states of a
two-dimensional (2D) Ising model reproduces the critical temperature of the
ferromagnetic phase transition, known from the exact solution~\cite{Onsager1944}. 
Following the seminal work, a variety of approaches are being explored to
test deep learning techniques in application to several models, including
the Ising and $q$-state Potts models, percolation, XY- and clock models~\cite{MorningstarMelko2018, Suchsland2018, Zhang2019XYpercolation,
Walker2020infogan, Fukushima2021, Miyajima2021, Shiina2020corr}.

It is becoming clear that a neural network (NN) trained on an equilibrium
ensemble of microscopic states can learn and predict phase transitions between
macroscopic states, \emph{in many situations.}
This gives rise to a series of fundamental questions: How to interpret NN
results from the physics perspective---specifically, does a NN learn the
critical behavior of a universality class of a transition? What are relevant NN
observables? How general is the NN approach and what are its failure modes? What
limits the reliability and
accuracy of these predictions? What is the role of the NN architecture?

In this Letter, we address these questions by considering two exactly solvable
models in 2D, the Ising model~\cite{Onsager1944} and the Baxter-Wu (BW) model~\cite{baxter1973exact,BaxterWu1974}.
We train NNs to perform binary classification of microscopic spin
configurations, and perform a careful finite-size scaling analysis of the 
classification results. We show that the second moment of the NN output displays
finite-size scaling governed by the correlation length exponent, $\nu$, of the
universality class of the model. We compare predictions of several
network architectures---fully connected networks (FCNN), shallow convolutional
networks (CNN) and several members of the ResNet family.

We note that using the BW model turns out to be essential to be able to
distinguish between the critical scaling, $\sim 1/L^\nu$, from regular,
analytic corrections, $\sim 1/L$, to thermodynamic limit behavior of systems
with finite linear size $L$. While for the Ising model the correlation length exponent $\nu=1$, the BW model
belongs to the 4-state Potts universality class with $\nu = 2/3$---thus making
the critical scaling clearly distinguishable from analytic corrections. We note that
the BW model, unlike other models in the same universality class, does not show any logarithmic corrections~\cite{baxter1973exact}, which allows us to simplify the finite-size analysis.

\textit{Models and methods.---} We consider two classical, exactly solved models,
formulated in terms of Ising spins, $\sigma_i = \pm 1$ on an $L\times L$ lattices.
The Ising model~\cite{Onsager1944} is defined by the Hamiltonian
$H_\mathrm{Is} = -J \sum_{\langle ij \rangle} \sigma_i \sigma_i$,
where $J$ is the coupling constant, and the summation runs over the pairs of
nearest neighbors of the \emph{square} lattice with periodic boundary conditions.
The BW model~\cite{baxter1973exact,BaxterWu1974} is defined on a triangular lattice, and contains three-spin interactions
$H_\mathrm{BW} = - J \sum_{\langle ijk \rangle} \sigma_i \sigma_j \sigma_k$,
where the summation runs over triplets of spins which form triangular plaquettes
of a triangular lattice with periodic boundary conditions. 
We consider the ferromagnetic case for both models and set $J=1$ for simplicity.

To generate data sets for NN training and validation, we use the standard Monte Carlo (MC) simulations
with Metropolis single spin flip updates~\cite{Metropolis1953}.
We use the Metropolis algorithm because we choose one modeling approach for two models: the Ising model and the Baxter-Wu model. It is known that the cluster algorithm~\cite{Novotny-1993} can lead to a shift of the cluster percolation from the critical point and thereby distort the critical behavior. At the same time, the Metropolis algorithm correctly reproduces the critical behavior of both models when taking into account the correlation time~\cite{Shchur-2010}.

 We perform simulations
for system sizes with $L=48, 72, 96, 144, \text{ and } 216$ for the Ising model, and
$L=48, 72, 96, 144, \text{ and } 243$ for the BW model. For each system size, we
perform simulations for $N_t = 114$ values of the temperatures between [$T_c - 0.4; T_c + 0.4$], using the value of the
critical temperature $T_c$ from the exact solution of the corresponding model.
For each system size and for each value of the temperature, we collect $N_s = 1500$
``snapshots'' of spin configurations generated by the MC process (here by a
``snapshot'' we mean a collection of $L^2$ spin values, $\pm 1$). To make sure
that snapshots are uncorrelated, we skip at least $2\,\tau_\mathrm{corr}$ Monte
Carlo steps between snapshots, where $\tau_\mathrm{corr}$ is the integrated autocorrelation time
for the magnetization~\cite{Sokal1997}. For each simulation, we allow at least
$20\,\tau_\mathrm{corr}$ MC steps for equilibration (see Ref.~\cite{chertenkov2022deep} for a detailed discussion of our MC simulations).

\textit{NN training.---} We train a NN to perform binary
classification of snapshots for a given system size $L$ into two classes,
ferromagnetic, FM, ($T < T_c$) or paramagnetic, PM, ($T > T_c$) separately for the Ising model and the BW model. 

A NN takes as input a ``snapshot'' of size $L\times L$, and outputs the class
scores for the FM and PM classes. We interpret the class scores as probabilities,
since their sum equal unity.

We use three different network architectures: Convolutional neural network
(CNN)~\cite{OShea2015CNN}, Fully-connected  neural network
(FCNN)~\cite{Schwing2015FCNN}, and Deep convolutional residual networks
(ResNet)~\cite{He2016deep}. In the ResNet family we use networks with 10, 18,
34, and 50 layers. Detailed parameters of the networks and our training protocol can be found in Supplemental Materials %
\footnote{See the Supplemental Material at [URL] for a detailed description of the NN architectures and training and inference protocols,
which includes Refs.\ \cite{
swendsen1987nonuniversal,wolff1989collective,mclib2022,Ratner2017NeurIPS,Kingma2014adam, Ruder2016adam}.}
.

\textit{Analysis of NN outputs.---} Once a NN is trained, we feed it with
$N$ snapshots from the testing dataset to perform the
classification. In what follows we denote by $f_i^T$ the FM class
prediction for the $i$-th snapshot at temperature $T$.

Averaging over the testing dataset, we define the average prediction, $F^T$,

\begin{equation}
F^T = \frac{1}{N} \sum_{i=1}^N f_i^T \\
\label{Pt}
\end{equation}

\noindent and its variance, $V^T$,

\begin{equation}
V^{T} =  \frac{1}{N}\sum_{i=1}^{N} \left(f_i^T \right)^2 -
        \left( \frac{1}{N}\sum_{i=1}^{N} f_i^T  \right)^2.
\label{Zt}
\end{equation}

\begin{figure*}[hbt]
  \centering
  \includegraphics[width=.9\columnwidth]{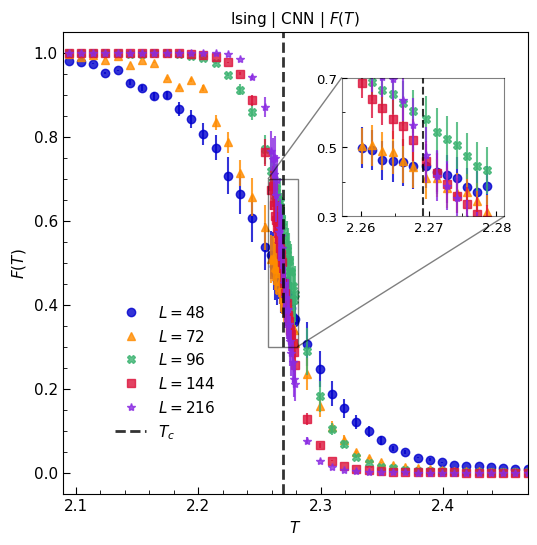} 
  \includegraphics[width=.9\columnwidth]{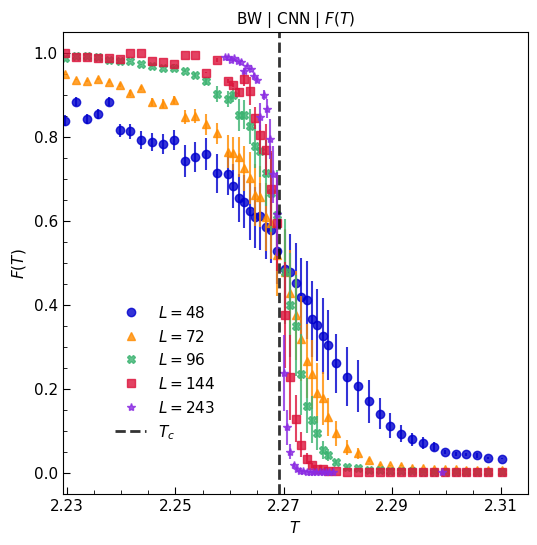} 
  \caption{$F^T$ ferromagnetic phase predictions for the Ising model (left) and the Baxter-Wu model (right) with FCNN for various lattice sizes.
The error bars correspond to the variance  $V^T$ of the NN prediction.  The black vertical dashed line is the position of the critical temperature  $T_c= 2/ \ln (1 + \sqrt{2})$, which by chance the same for both two models.
}
  \label{fig:Ft_conv}
\end{figure*}

Fig.~\ref{fig:Ft_conv} shows the dependence of the FM class prediction of (left image) the Ising model and (right image) the BW model with the CNN architecture. Other NN architectures give similar results. 
Here we only show the FM class prediction, because the PM class prediction is given by $1 - F^T$.

The network output, $F^T$, for both models, is clearly similar to the observation
of Ref.~\cite{Carrasquilla2017}: for low temperatures, $F^T \approx 1$, for
high temperatures, $F^T \approx 0$, and the transition region clearly shrinks
on increasing the system size $L$, thus developing a step function for $L\gg 1$.
This behavior is qualitatively similar for all network architectures we considered.

According to Ref.~\cite{Carrasquilla2017}---for the Ising model, the FM prediction, $F^T$,
approaches the value of 0.5 for all values of the system size $L$ at the exact value of the critical temperature, $T_c = 2/\ln(1 + \sqrt{2})$  \cite{Onsager1944}. Since the PM
prediction is simply $1 - F^T$, a straightforward interpretation would be that
at $T = T_c$, NNs are equally likely to classify a snapshot as either
ferromagnetic or paramagnetic \emph{for finite system sizes,} $L$.

However, our simulations of the Ising model and the BW model, Fig.\ \ref{fig:Ft_conv}, show that this interpretation is not entirely correct. 
For some lattice sizes for Ising model and for the BW model, the ``equal prediction''
point, $F^T = 1/2$ is shifted away from the value of
$T_c = 2/\ln(1 + \sqrt{2})$ known from the exact solution~\cite{baxter1973exact}. For FCNN architecture, the point $F^T = 1/2$ is shifted away to the paramagnetic phase for all lattice sizes both for the Ising and the BW models (see Fig.~2 of Supplemental Materials). Non-systematic shifts can be observed for the Ising and the BW models for different system sizes in the networks of the ResNet family. 
For the ResNet-50 (Fig.~6 of Supplemental Materials) for the Ising model large system sizes (96, 144, 216) are shifted to the ferromagnetic phase, while small sizes (48, 72) are shifted to the opposite side, to the paramagnetic phase.
We thus conclude that $F^T = 1/2$ is not a reliable finite-size estimate of the critical temperature $T_c$, in general. 

The average prediction $F^T = 1/2$ is correct for CNN applying the CNN to  the Ising model data, which is the case in Ref.~\cite{Carrasquilla2017}. We have found that this is generally not true for other networks, the ResNet family and FCNN. It probably depends on the technical parameters of the networks. Moreover, what we found that it does not apply in general to other models of statistical mechanics. This is probably due to the symmetry of the ground state of the models. It is well known 2D Ising model have many hidden symmetries, and care should be taken to transfer knowledge from the Ising model and apply it the other models.

For the Ising model, Ref.~\cite{Carrasquilla2017}, considered system sizes of
up to $L=60$ and observed that the $F^T$ curves display data collapse with
respect to the ``scaling variable'', $t L^{1/\nu}$, where the reduced
temperature $t = (T - T_c) / T$ is scaled by the critical exponent $\nu$. The
data collapse estimate of Ref.~\cite{Carrasquilla2017} for $L$ up to $L=60$,
produces the values $T_c = 2.266 \pm 0.002$ and $\nu = 1.0 \pm 0.2$, consistent
with the exact values of the critical temperature and the 
correlation length exponent for the 2D Ising universality class, $\nu{=}1$, \cite{baxter1973exact}.
Our numerical experiments show that data collapse is visually observed in a wide range of values of the critical exponent $\nu \in [0.75, 1.5]$, depending on the network architecture (see Figs 12-23 of Supplemental Materials for details). We stress that simply including larger system sizes does not improve correlation length exponent and critical temperature
estimates due to increasing errorbars of the NN output in the critical region, cf Fig.\ref{fig:Ft_conv}.

We note however, that the increase of the errobars of $F^T$, Eq.\ \eqref{Pt}---equivalently,
the variance $V^T$, Eq.\ \eqref{Zt}---around $T=T_c$ is similar to the expected
behavior of thermodynamic functions in the critical region, where second moments
of observables are related to temperature derivatives of corresponding thermodynamic
functions. In this spirit, we consider the second moment of the NN prediction of the FM class,
Eq.~\eqref{Zt}, and hypothesize that the variance of the NN output, Eq.\ \eqref{Zt}
is singular in the thermodynamic limit. This way, the observed increase of the errorbars of $F^T$ around $T \approx T_c$ is in fact nothing but a finite-size rounding of this divergence, governed by the
correlation length exponent $\nu$.
Incremental cutoff values are applied to low values of $V(T)$ and $T$ range until the Gaussian fit parameters become stable. The optimal parameters $p_{opt}$ are obtained by minimizing the square deviation $V^T_{fit} - V^T$ of the non-linear least squares method. With the parameters $p_{opt}$, we estimate the standard deviation $p_{sd}$, which is obtained as a linear approximation of the model function around the optimum~\cite{vugrin-2007}. 
We  used the built-in functions of the \textit{scipy} package~\cite{scipy-2020} to get $p_{opt}$, $p_{sd}$.

\begin{figure*}[hbt]
  \centering
  \includegraphics[width=.9\columnwidth]{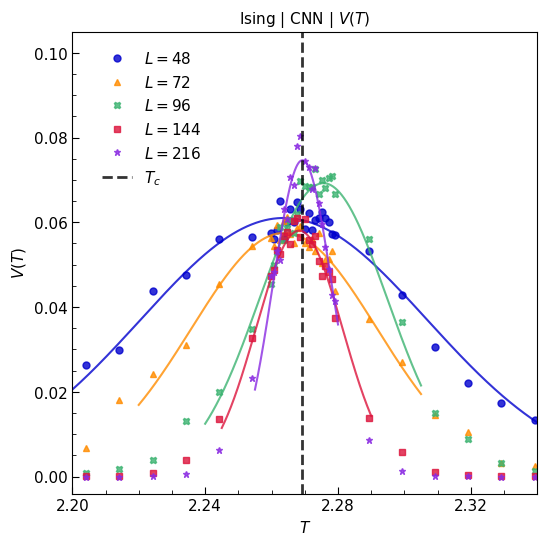}
  \includegraphics[width=.9\columnwidth]{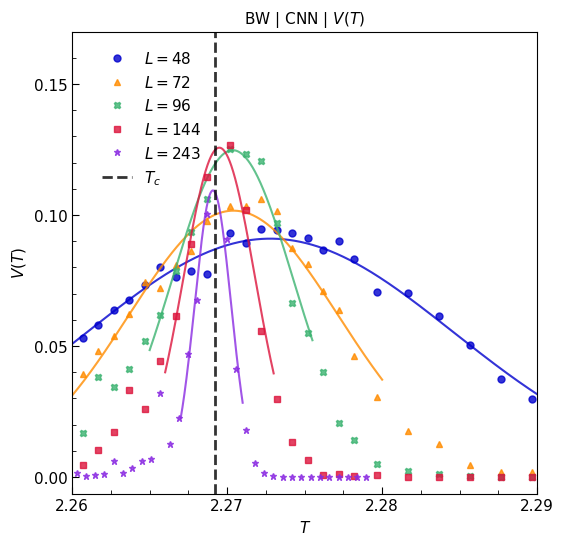} 
  \caption{$V^T$ variance for the Ising model (left) and the Baxter-Wu model (right) with FCNN for different lattice sizes. The black vertical dahed line is the position of the critical temperature. The solid lines are limited to the area where the Gaussian approximation was applied to extract the width $\sigma$ for each lattice size $L$.}
  \label{fig:Vt_conv}
\end{figure*}

Fig.~\ref{fig:Vt_conv} displays the temperature dependence of $V^T$, which
indeed shows a drastic increase around $T=T_c$, and a characteristic
Gaussian-like bell shape for both Ising and BW models and all network
architectures. Furthermore, the widths of the bell-shaped curves decrease with
increasing the system size, which is consistent with scaling behavior. 

To test this hypothesis, we study the $L$ dependence of the width of the peak
of $V^T$, Eq.\ \eqref{Zt}. Specifically, for each value of $L$, we fit $V^T$ vs T with an
unnormalized Gaussian-like Ansatz, $V^T \sim \exp\left(-(T-T_*)^2 / 2\sigma^2\right)$,
with $\sigma$ and $T_*$ being fit parameters, 
and extract the dependence of the width of $\sigma$ on $L$. Since there is no
\textit{a priori} requirement that the profile is strictly Gaussian, we also
perform a separate single-parameter fits the left-hand ($T < T_*$) and the right-hand ($T > T_*$)
parts of the $V^T$ curves. In this procedure, $T_*$ is simply the location of the maximum of
$V^T$, and $\sigma$ is the (only) fit parameter. 
For both fitting protocols, we then fit the resulting widths, $\sigma(L)$, to a power-law Ansatz,
$\sigma(L) \sim 1/L^{1/\nu_\sigma}$. 
Similarly to the Gaussian fitting, we obtain the optimal value of $1/\nu_{\sigma}$ and its standard deviation from the power-law fitting.
We perform this procedure for the Ising
and the BW models and for all network architectures, and results are summarized
in Tables \ref{tab-IS} and \ref{tab-BW}.

For the Ising model, Table \ref{tab-IS}, the first observation is that
the resulting values of the scaling exponent (both one-sided $1/\nu_{\sigma}$
 and two-sided $1/\nu_{\sigma^\pm}$) are consistent with the
correlation length exponent for the Ising universality class, $\nu = 1$. 
One notable exception is the ResNet 10- and 34-layer architectures, which shows
vastly different values for exponents $1/\nu_{\sigma^{\pm}}$ and $1/\nu_{\sigma}$,
and the resulting values are barely within the 4 standard deviations from the
exact result, $\nu=1$.

For the BW model, Table \ref{tab-BW}, the striking observation is that the
scaling exponents, $1/\nu_{\sigma}$, estimated from the width of $V^T$, are
consistent with the exact value of the correlation length exponent for
universality class of the BW model, $\nu = 2/3$. The accuracy of fit results, Table \ref{tab-BW},
allows to conclusively distinguish this value from regular, non-singular corrections, $\sim L^{-1}$.
This is the major advantage of considering the BW model in addition to the Ising
model where $\nu=1$.

We also note that the shape of $V^T$ is in fact not symmetric around the maximum---for both Ising and BW models. Allowing for different widths, $\sigma^+$ and $\sigma^-$ for $T > T_*$ and $T < T_*$,
respectively, produces closer fits of $V^T$. Moreover, scaling exponents, $1/\nu_{\sigma^+}$ and $1/\nu_{\sigma^-}$ are different---the low-temperature exponent, $1/\nu_{\sigma^-}$, is consistently larger than the high-temperature
exponent, $1/\nu_{\sigma^+}$ --- again, for both Ising and BW models.

It is clear from Tables \ref{tab-IS} and \ref{tab-BW} that the values of the critical exponents, extracted from NN data are largely independent of the NN architecture, and that increasing the depth of an
NN does not bring drastic improvements in exponent accuracy estimation. For networks of the ResNet family, both for the Ising model and for the BW model, some of the scaling exponents have larger errors than similar ones for simpler architectures FCNN and CNN. 

We thus conclude that the width of the $V^T$ peak displays finite-size scaling consistent with the universality class of a model, and that simple convolutional networks, CNN, or fully-connected, FCNN, are
more appropriate for studying this class of problems, and that increasing the
network depth does not automatically translate into better reliability or
accuracy of the estimates---this is consistent with the conclusion of
Ref.~\cite{MorningstarMelko2018}.

Given that the width of the $V^T$ peak displays finite size scaling with the correlation length exponent, it is natural to study $L$-dependence of other properties of the peak: its maximum value, $V_\mathrm{max}^T$, and the shift of the maximum from the thermodynamic limit value of $T_c$.
Our numerical experiments show that both maximum height and the peak shift are NN architecture dependent and do not display meaningful convergence with $L\to\infty$.

This behavior must be contrasted with the behavior of more traditional thermodynamic observables. It is well known~\cite{Ferdinand1969} that the position of the specific heat maximum $T^*$ shifts from the critical point $T_C$ with the correlation length index $T^*-T_C \propto 1/L^{1 /\nu}$, and the same behavior is found for other thermodynamic quantities due to fluctuation cutoff, when the correlation length becomes comparable with the dimensions of the system, similar to i.e. to the rounding of the magnetic susceptibility at a temperature close to the critical one~\cite{Landau2014}. 

We tested the deviation of the maximum VOT for both models and six networks, and the results are placed in Table~\ref{tab-IS-TC} for the Ising model and Table~\ref{tab-BW-TC} for the Baxter-Wu model. Note that the critical temperature values are coincidentally the same for the two models, but $1/\nu$ is different -- it is 1 for the Ising model and 1.5 for the Baxter-Wu model, and we use these values when analyzing the VOT data. A demonstration of fits can be found in Supplemental Materials. The results of the fitting are in most cases consistent within no more than five standard deviations and follow the assumption that the shift of the VOT function follows the Ferdinand-Fischer law with an exactly known exponent. The testing of the Ising model with the ResNet-50 network is the worst, and at the same time the $T*$ estimates for the largest systems are very close to the critical temperature $T_C$, as can be seen from the Fig.~23 in Supplemental materials. Surprisingly, the values of $T*$ change more regularly with $L^{-1/\nu}$ for the Baxter-Wu model than for the Ising model. This may be due to weaker corrections to scaling for the Baxter-Wu model (see, for discussion Ref.~\cite{Shchur-2010}).

\begin{table}[h!]
\centering
\begin{tabular}{|l|r|r|r|} 
\hline\hline
NN & $1/\nu_\sigma$  &  $1/\nu_{\sigma^-}$ & $1/\nu_{\sigma^+}$ \\ \hline
FCNN & 1.01(1) & 1.02(13) & 0.98(4) \\
CNN & 1.06(3) & 1.11(5) & 1.07(2)  \\ 
ResNet-10 & 1.25(3) & 1.24(7) & 1.24(3) \\
ResNet-18 & 1.17(11) & 1.41(6) & 1.08(10) \\
ResNet-34 & 1.15(16) & 1.26(7) & 1.12(24) \\
ResNet-50 & 1.20(5) & 1.21(5) & 1.31(6) \\
\hline\hline
\end{tabular}
\caption{Peak widths for the Ising model. Here $\nu_\sigma$ is the estimate
from fitting the Gaussian profile to the $V^T$. $\nu_\sigma^+$
and $\nu_\sigma^-$ are similar estimates where we only fit the right-hand side
(resp., left-hand-side) of the $V^T$ curves. See the text for discussion.}
\label{tab-IS}
\end{table}

\begin{table}[h!]
\centering
\begin{tabular}{|l|r|r|r|} 
\hline\hline
NN & $1/\nu_\sigma$  &  $1/\nu_{\sigma^-}$ & $1/\nu_{\sigma^+}$  \\ \hline
FCNN & 1.49(3) & 1.57(2) & 1.38(8) \\
CNN & 1.45(5) & 1.55(6) & 1.49(5) \\ 
ResNet-10 & 1.48(5) & 1.65(13) & 1.47(4) \\
ResNet-18 & 1.32(11) & 1.36(14) & 1.40(7) \\
ResNet-34 & 1.54(6) & 1.76(5) & 1.47(3) \\
ResNet-50 & 1.43(9) & 1.69(16) & 1.47(5) \\
\hline\hline
\end{tabular}
\caption{Peak widths for the Baxter-Wu model. Here $\nu_\sigma$, $\nu_\sigma^+$,
and $\nu_\sigma^-$ are the same as in Table~\ref{tab-IS}.}
\label{tab-BW}
\end{table}

\begin{table}[h!]
\centering
\begin{tabular}{|l|l|r|} 
\hline\hline
NN & $T*$  &  $\Delta/\sigma_T$   \\ \hline
FCNN & 2.2699(5) & 1  \\
CNN & 2.2727(6) & 5  \\ 
ResNet-10 & 2.2667(6) & 4.2  \\
ResNet-18 & 2.2688(6) & 0.7  \\
ResNet-34 & 2.2659(6) & 5.5 \\
ResNet-50 & - & -  \\
\hline\hline
\end{tabular}
\caption{Ising model: estimation of the critical temperature from the VOT width using Ferdinand-Fisher law. The last column is the difference between the estimated critical temperature and the exact critical temperature $\Delta=|T^*-T_c|$ divided by the statistical error $\sigma_T$ of the weighted linear fit.}
\label{tab-IS-TC}
\end{table}

\begin{table}[h!]
\centering
\begin{tabular}{|l|l|r|} 
\hline\hline
NN & $T*$  &  $\Delta/\sigma_T$   \\ \hline
FCNN & 2.2691(4) & 0  \\
CNN & 2.2687(4) & 1.25  \\
ResNet-10 & 2.2690(4) & 0.25  \\
ResNet-18 & 2.2684(4) & 2  \\
ResNet-34 & 2.2694(4) & 0.5  \\
ResNet-50 & 2.2688(4) & 1 \\
\hline\hline
\end{tabular}
\caption{Baxter-Wu model: estimation of the critical temperature from the VOT width using Ferdinand-Fisher law. The last column is the difference between the estimated critical temperature and the exact critical temperature $\Delta=|T^*-T_c|$ divided by the statistical error $\sigma_T$ of the weighted linear fit.}
\label{tab-BW-TC}
\end{table}

\textit{Conclusion.---} The main result of the presented analysis is that the most reliable information on the classification of snapshots of the spin configuration of statistical mechanics systems experiencing phase transitions of the second kind is contained in the output variation (VOT) of neural networks. Namely, VOT contains information about the critical temperature and the correlation length exponent. We present a VOT analysis method and extract estimates for the critical temperature and correlation length exponent of two systems in two universality classes. The results are stable when using three different architectures in the NN deep pool - CNN, FCNN and Resnet with four configurations.

We do not have theory for the network output function as the thermodynamic function in the same ensemble as the statistical mechanics model which we tested with the neural network. At the same time we found evidence that the VOT width scales with the critical length exponent $\nu$ and demonstrated that clearly for two universality classes. This means that output function $F(T)$ somehow connected to the fluctuation of the physical quantities of the model although the clear connection is not directly found~\footnote{It should be noted an example in which the numerical detection of giant deviations of thermodynamic quantities in the critical region due to the impact of a random number generator (RNG)~\cite{Ferrenberg1992} was subsequently explained~\cite{Shchur1997}
as a resonance of the RNG shift register length with the cluster size -- due to the scaling of the Wolf cluster size, which is the magnetic susceptibility at the critical temperature, the deviations are also scales in the critical region with {\em some exponents}, and the corresponding width follows the Ferdinand-Fisher law.}. 

We find no evidence that the network output function $F(T)$ should be equal to 1/2 at the critical point, as stated in the pioneering work~\cite{Carrasquilla2017} --- our claim is based on careful analysis using different network architecture. Instead, we show that the variation bias of the VOT output function does not contradict the Ferdinand-Fischer picture and can be used to estimate the critical temperature. This estimate is still not under control of the desired accuracy and more work needs to be done on a sound methodology.

We would like to emphasize that the width dependence of VOT on the system size is a good candidate for extracting the exponent $\nu$ of the critical length and gives better accuracy than the approach proposed in Ref.~\cite{Carrasquilla2017} using $F(T)$ collapse data. We should note again that more research is needed to find a reliable way to estimate $\nu$ from the VOT width, since not all network architectures produce $\nu$ with the desired precision.

\begin{acknowledgments}
Research supported by the grant 22-11-00259 of the Russian Science Foundation. 
The simulations were done using the computational resources of HPC facilities at HSE University~\cite{kostenetskiy2021hpc}.
\end{acknowledgments}

\bibliography{sources}

\end{document}